\documentclass[twocolumn,superscriptaddress,prl,showpacs]{revtex4}
\usepackage{amsmath}
\usepackage{amssymb}
\usepackage{graphicx}
\usepackage{bm}

\begin{document}

\title{Thermal Casimir-Polder shifts in Rydberg atoms near metallic surfaces}

\author{J. A. Crosse}
\email{jac00@imperial.ac.uk}
\affiliation{Quantum Optics and Laser Science, Blackett Laboratory, 
Imperial College London, Prince Consort Road, London SW7 2AZ}
\author{Simen \AA. Ellingsen}
\affiliation{Department of Energy and Process Engineering, Norwegian
University of Science and Technology, N-7491 Trondheim, Norway}
\author{Kate Clements}
\author{Stefan Y. Buhmann}
\author{Stefan Scheel}
\affiliation{Quantum Optics and Laser Science, Blackett Laboratory, 
Imperial College London, Prince Consort Road, London SW7 2AZ}

\date{\today}

\begin{abstract}
The Casimir-Polder (CP) potential and transition rates of a Rydberg atom above
a plane metal surface at finite temperature are discussed. As an example, the
CP potential and transition rates of a rubidium atom above a copper surface at 
300k are computed. Close to the surface we show that the quadrupole
correction to the force is significant and increases with increasing principal
quantum number $n$. For both the CP potential and decay rates one finds that the
dominant contribution comes from the longest wavelength transition and the
potential is independent of temperature. We provide explicit scaling laws for 
potential and decay rates as functions of atom-surface distance and principal
quantum number of the initial Rydberg state.
\end{abstract}


\maketitle
%%%%%%%%%%%%%%%%%%%%%%%%%%%%%%%%%%%%%%%%%%%%%%%%%%%%%%%%%%%%%%%%%%%%%%

Rydberg atoms -- atoms excited to large principal quantum numbers $n$ -- have
attracted much attention in recent decades \cite{gallagher}. Aside from the
inherent interest of studying such extreme states, the exaggerated properties of
these highly excited atoms make them ideal for examining the properties of a
variety of systems that would be awkward to probe by other means. The large
cross-sections and weakly bound outer electrons associated with Rydberg atoms
makes them extremely sensitive to small-scale perturbations and dispersion
potentials, such as the van der Waals (vdW) and Casimir--Polder (CP)
potentials \cite{casimir}.

For example, the strong scaling of the free-space vdW potential between two
Rydberg atoms with $n$  ($\propto n^{11}$) leads to the Rydberg blockade
mechanism which has been put forward as a candidate for
implementing controlled gate operations between isolated atoms
\cite{Lukin,Pfau}. The effect relies on the massive level shift that one Rydberg
atom experiences in close proximity to another. 

Level shifts of similar origin arise if the atoms are brought into the
vicinity of a macroscopic body. With the increasing ability to trap and
manipulate atoms close to macroscopic bodies, the effects of these surface (CP)
potentials have become a subject of great interest. Applications ranging from
novel atom trapping methods \cite{prism} to atom chip physics \cite{atomchip}. 
Thus, it is of both fundamental and practical interest to understand the interplay 
between atoms in highly excited states and field fluctuations emanating from 
macroscopic bodies.

In this Letter we provide evidence that dispersion forces have a sizeable effect
on the energy levels of highly excited Rydberg atoms when brought close to
metallic surfaces, with shifts on the order of several GHz expected at
micrometer distances. Due to the large atom size, next-to-leading order terms in
the multipole expansion of the radiation field give additional contributions in
the MHz range. Despite the existence of large numbers of thermal photons at 300K
at the relevant atomic transition frequencies, the level shifts are in fact 
temperature-independent \cite{tinvariance}.

For a given atom-field coupling $\hat{H}_\mathrm{int}$, the CP potential for an
atom in state $|n\rangle$ and the radiation field in state $|q\rangle$ is
given by the position-dependent part of the energy shift which, to second order
in perturbation theory, reads 
\begin{equation}
\delta E_n = \langle n,q|\hat{H}_\mathrm{int}|n,q\rangle \\
 +\sum\limits_{n',q' \neq n,q}
\frac{|\langle n,q|\hat{H}_\mathrm{int}|n',q'\rangle|^{2}}
{E_{n+q} -E_{n'+q'}} 
\label{dw}
\end{equation}
where $E_{n+q}$ are the unperturbed energy eigenvalues of the atom-field system.
In the long-wavelength approximation, the electric field couples to the atomic
dipole moment $\hat{\mathbf{d}}$ via the interaction Hamiltonian
\begin{equation}
\hat{H}_\mathrm{int} = -\hat{\mathbf{d}}\cdot\hat{\mathbf{E}}(\mathbf{r}_{A}),
\label{dHint}
\end{equation}
with the electric field given in terms of the classical Green tensor (for a
recent review see, e.g.~\cite{slovaca})
\begin{equation}
\hat{\mathbf{E}}(\mathbf{r}) = \displaystyle\sum\limits_{\lambda = e,m} \int
d^3r'\int d\omega\,\bm{G}_{\lambda}(\mathbf{r},
\mathbf{r}',\omega)\cdot\hat{\mathbf{f}}_{\lambda}(\mathbf{r}',\omega) +
\mbox{h.c.}, 
\label{E}
\end{equation}
with
\begin{align}
&\bm{G}_{e}(\mathbf{r}, \mathbf{r}', \omega) =
i\frac{\omega^{2}}{c^{2}}\sqrt{\frac{\hbar}{\pi\varepsilon_{0}}
\mathrm{Im}\varepsilon(\mathbf{r}',\omega)}
\bm{G}(\mathbf{r}, \mathbf{r}', \omega)\,,\\ 
&\bm{G}_{m}(\mathbf{r}, \mathbf{r}', \omega) =
-i\frac{\omega}{c}\sqrt{\frac{\hbar}{\pi\varepsilon_{0}}
\frac{\mathrm{Im}\mu(\mathbf{r}',\omega)}{|\mu(\mathbf{r}',\omega)|^2}}
\left[ \bm{G}(\mathbf{r},\mathbf{r}',\omega) \times
\overleftarrow{\bm{\nabla}}' \right].
\end{align}
The Green tensor $\bm{G}(\mathbf{r}, \mathbf{r}', \omega)$ solves the Helmholtz
equation for a point source and contains all the information about the geometry
of the system. The bosonic vector fields
$\hat{\mathbf{f}}_\lambda(\mathbf{r},\omega)$
describe collective excitations of the electromagnetic field and the linearly
absorbing dielectric matter.

The CP potential at temperature $T$ acting on an atom in
state $|n\rangle$ via a dipole interaction~\eqref{dHint} is given by
\cite{thermalcp}
\begin{gather}
U_\mathrm{CP}^\mathrm{dip}(\mathbf{r}_A) = \mu_{0}k_BT\sum\limits_{j=0}^\infty
\!{}' \xi_{j}^{2}
\bigg[\bm{\alpha}(i\xi_{j})\bullet\bm{G}^{(1)}(\mathbf{r}_{A},\mathbf {r}_{A},
i\xi_{j})\bigg]
\nonumber\\
+ \mu_{0}\sum_{k \neq n}\omega_{kn}^{2}n(\omega_{kn})\,
\big(\mathbf{d}_{nk}\otimes\mathbf{d}_{kn}
\big)\bullet\mathrm{Re}\bm{G}^{(1)}(\mathbf{r}_{A}, \mathbf{r}_{A},\omega_{kn})
\label{cpd}
\end{gather}
where $\bullet$ denotes the Frobenius inner product
($\bm{A}\bullet\bm{B}=\sum_{i_1\ldots i_k}A_{i_1\ldots i_k}B_{i_1\ldots i_k}$) 
and the primed summation means that the term with $j=0$ contributes only with
half weight. Here $\bm{G}^{(1)}(\mathbf{r}_{A}, \mathbf{r}_{A},\omega)$ is the
scattering part of the Green tensor. The atomic polarizability is defined as
\begin{equation}
\bm{\alpha}(\omega) = \bigg[\frac{1}{\hbar}
\displaystyle\sum_{k \neq n}\frac{\mathbf{d}_{nk}\otimes
\mathbf{d}_{kn}}{(\omega_{kn} + \omega)} +
\frac{\mathbf{d}_{nk}\otimes\mathbf{d}_{kn}}{(\omega_{kn} - \omega)}\bigg],
\end{equation}
with $\omega_{kn}=(E_{k}-E_{n})/\hbar$ denoting the atomic transition
frequencies. The frequencies $\xi_{j}=2\pi k_BTj/\hbar$, $j\in\mathbb{N}$ 
are the Matsubara frequencies and $n(\omega)=[e^{\hbar\omega/k_BT}-1]^{-1}$ is
the thermal photon number distribution.

The reflective part of the scattering Green tensor of an infinitely extended planar metal that fills
the lower half space $z<0$ is given by \cite{chew}
\begin{equation}
\bm{G}^{(1)}(\mathbf{r},\mathbf{r}',\omega) =
\int\frac{d^2k_\|}{(2\pi)^2}\bm{R}(\mathbf{k}_\|,z,z',\omega)
e^{i\mathbf{k}_\|\cdot(\mathbf{r}_\|- \mathbf{r}'_\|)},
\label{G}
\end{equation}
with $\mathbf{r}_\| = (x,y,0)$, $\mathbf{k}_\| = (k_{x},k_{y},0)$ and
$k_\|=|\mathbf{k}_\||$. The reflection tensor
$\bm{R}(\mathbf{k}_\|,z,z',\omega)$ has the form
\begin{align}
\bm{R}(\mathbf{k}_\|,z,z',\omega) =
\frac{-i}{8\pi^2\beta_{+}}\sum\limits_{\sigma=s,p}
r_{\sigma}e^{i\beta_{+}(z+z')} 
\mathbf{e}^{+}_{\sigma}\otimes\mathbf{e}^{-}_{\sigma}.
\end{align}
Here the unit vectors for $s$-polarized and $p$-polarized waves are
$\mathbf{e}^{\pm}_{s} = \mathbf{e}_{k_\|}\times\mathbf{e}_{z}$ and
$\mathbf{e}^{\pm}_{p}=(k_\|\mathbf{e}_{z}\mp\beta_{+}\mathbf{e}_{k_\|})/q$.
The functions 
$r_s=[\varepsilon(\omega)\beta_+-\beta_-]/[\varepsilon(\omega)\beta_++\beta_-]$
and $r_p=[\beta_+-\beta_-]/[\beta_++\beta_-]$ are the usual Fresnel reflection
coefficients for those waves with wave numbers
$\beta_-=\sqrt{q^2\varepsilon(\omega)-k_\|^2}$ and
$\beta_+=\sqrt{q^2-k_\|^2}$, and $q=\omega/c$. 
The permittivity of the metal surface is modelled by the Drude relation,
$\varepsilon(\omega) = 1-\omega^{2}_\mathrm{p}/\omega(\omega + i\gamma)$,
where $\omega_\mathrm{p}$ and $\gamma$ are the plasma frequency and the
relaxation rate of the metal, respectively. Magnetic effects will be neglected.

Matrix elements of the dipole operator
$\hat{\mathbf{d}}=e\hat{\mathbf{r}}=e\hat{r}\mathbf{e}_{r}$
for the transition between two electronic states $|n,l,j,m\rangle$ 
($n$, principal quantum number; $l$, $j$, $m$, quantum numbers for orbital and
total angular momentum and $z$-component of the latter) and 
$|n',l',j',m'\rangle$ factor into a radial and an angular part according to
\begin{multline}
\langle n',l',j',m'| \hat{\mathbf{d}} |n,l,j,m\rangle\\ 
= e \langle R_{n',l',j'}| \hat{r} |R_{n,l,j}\rangle
\langle l'j'm'| \mathbf{e}_{r} |ljm\rangle,
\label{dl}
\end{multline}
where $|R_{n,l,j}\rangle$ are the radial wavefunctions. The radial matrix
elements are computed numerically using the Numerov method
\cite{numerov1,numerov2} in which the suitably scaled radial Schr\"odinger
equation is integrated inwards until an inner cut-off point (commonly the radius
of the rump ion). The eigenenergies are computed as
$E_{n,l,j}=-\mathrm{Ry}/n^{\ast 2}$ (Ry: Rydberg constant) where
$n^\ast=n-\delta_{n,l,j}$ is the effective quantum number and $\delta_{n,l,j}$
the quantum defect \cite{quantumdefect} whose values are tabulated in the
literature \cite{Li03}.

To evaluate the angular part, we first convert from the $jm$ basis to a
$m_{l}m_{s}$ basis ($m_{l}$, $m_{s}$: $z$-components of orbital angular momentum
and spin) by summing over the relevant Clebsch-Gordan coefficients,
\begin{multline}
\langle l'j'm'|\mathbf{e}_{r}|ljm\rangle 
=\displaystyle\sum_{\substack{m_{l}m'_{l}\\m_{s}}}
C^{j,l,1/2}_{m,m_{l},m_{s}}C^{j',l',1/2}_{m',m'_{l},m_{s}}
\\[-3ex]
\times \langle Y_{l',m'_{l}}|\mathbf{e}_{r}|Y_{l,m_{l}}\rangle .
\label{dj}
\end{multline}
with the orbital-angular momentum eigenstates $|Y_{l,m_{l}}\rangle$
being spherical harmonics. Matrix elements in the $m_{l}m_{s}$ basis are computed by
rewriting the radial unit vector in terms of spherical harmonics 
[$Y_{lm}\equiv Y_{lm}(\vartheta,\varphi)$]
\begin{equation}
\mathbf{e}_{r} = 
\sqrt{\frac{2\pi}{3}}\left(\begin {array}{c} Y_{1,-1} - Y_{1,1} \\ 
i\left(Y_{1,-1} + Y_{1,1}\right) \\ 
\sqrt{2}Y_{1,0} \end{array}\right).
\label{yd}
\end{equation}
and using the integral relation [$d\Omega\equiv
\sin\vartheta\,d\vartheta\,d\varphi$]
\begin{gather}
\int d\Omega\,Y_{l_{1},m_{1}}Y_{l_{2},m_{2}}Y_{l_{3},m_{3}}
%=\sqrt{\frac{\left(2l_{1}+1\right)\left(2l_{2}+1\right)
%\left(2l_{3}+1\right)}{4\pi}}  \nonumber\\
=\sqrt{{\textstyle\frac1{4\pi}\prod_{\nu=1}^3}(2l_\nu+1)}  \nonumber\\
\times
\left(\begin{array}{ccc} l_{1} & l_{2} & l_{3} \\ 0 & 0 & 0 \\
\end{array}\right)
\left(\begin{array}{ccc} l_{1} & l_{2} & l_{3} \\ m_{l_{1}} &
m_{l_{2}} & m_{l_{3}} \\ \end{array}\right)
\label{inteq}
\end{gather}
that expresses the angular integral over three spherical harmonics in
terms of Wigner 3j-symbols.

When Rydberg atoms are held sufficiently close to a 
% metallic 
surface, its effective radius $\langle r\rangle\simeq a_0n^2$ [$a_0$: Bohr
radius] can be on the order of micrometres and therefore a significant fraction
of the surface distance. The dipole approximation is then no longer appropriate.
In other words, the atom cannot be viewed as a point-like particle, and its
non-negligible size requires the inclusion of contributions from higher-order
multipoles. This correction can be found via a similar method as described
above, with the dipole interaction Hamiltonian replaced by the quadrupole
interaction Hamiltonian \cite{decay1}
\begin{equation}
\hat{H}_\mathrm{int} = -\hat{\mathbf{Q}}\bullet\left[\bm{\nabla}\otimes
\hat{\mathbf{E}}(\mathbf{r}_{A})\right].
\label{QHint}
\end{equation}
In close analogy to the dipole case the CP potential for a quadrupole
interaction is found to be
\begin{align}
&U_\mathrm{CP}^\mathrm{quad}(\mathbf{r}_{A}) = 
\mu_{0}k_BT\sum\limits_{j=0}^\infty \!{}' \xi_{j}^{2}
\bm{\alpha}^{(4)} (i\xi_{j})\notag \\
&\bullet\bigg[
\bm{\nabla}\otimes
\bm{G}(\mathbf{r}_{A},\mathbf{r}_{A},i\xi_{j})\otimes\overleftarrow{
\bm{\nabla}}
\bigg]
+ \mu_{0}\sum\limits_{k \neq n}\omega_{kn}^{2}n(\omega_{kn})
\notag\\
&\times\big(\mathbf{Q}_{nk}\otimes\mathbf{Q}_{kn}\big) 
\bullet \big[\bm{\nabla}\otimes \mathrm{Re}\bm{G}(\mathbf{r}_{A},
\mathbf{r}_{A},\omega_{kn})\otimes\overleftarrow{\bm{\nabla}}\big],
\label{cpq}
\end{align}
with the quadrupole moment operator
$\hat{\mathbf{Q}}=e(\hat{\mathbf{r}}\otimes\hat{\mathbf{r}})/2$ and the atomic
quadrupole polarizability defined as
\begin{equation}
\bm{\alpha}^{(4)}(\omega) = \frac{1}{\hbar}\displaystyle\sum_{k \neq
n}\bigg[\frac{\mathbf{Q}_{nk}\otimes\mathbf{Q}_{kn}}{(\omega_{kn} +
\omega)} +
\frac{\mathbf{Q}_{nk}\otimes\mathbf{Q}_{kn}}{(\omega_{kn} - \omega)}\bigg].
\end{equation}
The matrix elements for the quadrupole transitions can again be evaluated by
factoring 
$\hat{\mathbf{Q}}=(e/2)\hat{r}^2
\mathbf{e}_r\otimes\mathbf{e}_r$
and computing the matrix elements between the radial and angular parts
of the wave functions separately. Evaluation of the radial integral is
again performed numerically. The tensor product of unit vectors in
spherical harmonic form reads
$\mathbf{e}_r\otimes\mathbf{e}_r=\sqrt{\frac{2\pi}{15}}\mathbf{A}$ with
\begin{subequations}
\begin{align}
A_{{}^{xx}_{yy}}=&\pm Y_{2,-2} \pm Y_{2,2} -
\sqrt{{\textstyle\frac{2}{3}}}Y_{2,0} +
\sqrt{{\textstyle\frac{10}{3}}}Y_{0,0},\\
A_{xy}=&A_{yx} = i\left(Y_{2,-2} -Y_{2,2}\right),\\
A_{xz}=&A_{zx} = Y_{2,-1} - Y_{2,1},\\
A_{yz}=&A_{zy} = i\left(Y_{2,-1} + Y_{2,1}\right),\\
A_{zz}=&\sqrt{{\textstyle\frac{8}{3}}}Y_{2,0} +
\sqrt{{\textstyle\frac{10}{3}}}Y_{0,0}.
\end{align}
\end{subequations}
The angular matrix elements can then be evaluated using Eq. \eqref{inteq}.

\begin{figure}[tb]
\includegraphics[width=8cm]{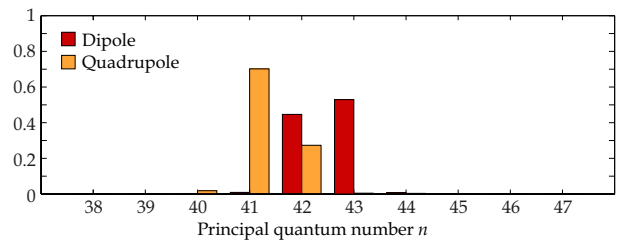}
\caption{Relative contributions from different transitions to the CP dipole
% (dark) 
and quadrupole 
% (light) 
level shift of the state $43s$ of ${}^{87}$Rb.} 
\label{fig:transitions}
\end{figure}

As can be seen from Eqs.~\eqref{cpd} and \eqref{cpq}, the CP potential is
comprised of a pair of sums, one over the Matsubara frequencies and one over
all available atomic transitions. It turns out, however, that due to the
finite-temperature environment only a limited number of dipole 
%SS
and quadrupole
transitions contribute significantly to the total level shift. This effect is
depicted in Fig.~\ref{fig:transitions}, where we show the relative contributions
of the dipole transitions $43s\to np$ and quadrupole transitions $43s\to nd$ to
the total level shift of the $43s$ state of ${}^{87}$Rb. Note that the final $n$ 
state for the dominant dipole and quadrupole transition is shifted. This is due to the differing 
quantum defects for the $p$ and $d$ states. Moreover, for each of
these individual (long-wavelength) 
% dipole 
transitions the first term in the
Matsubara sum (with $j=0$) dominates at the micrometer atom-surface distances
envisaged here, and all other terms can be safely neglected. Remarkably, we
observe that the CP potential is independent of temperature from $T=0$ to $300$K
and beyond. As was recently shown \cite{tinvariance}, this is due to the
dominance of contributions from transitions whose wavelengths far exceed
atom-surface separations.

\begin{figure}[tb]
\includegraphics[width=8cm]{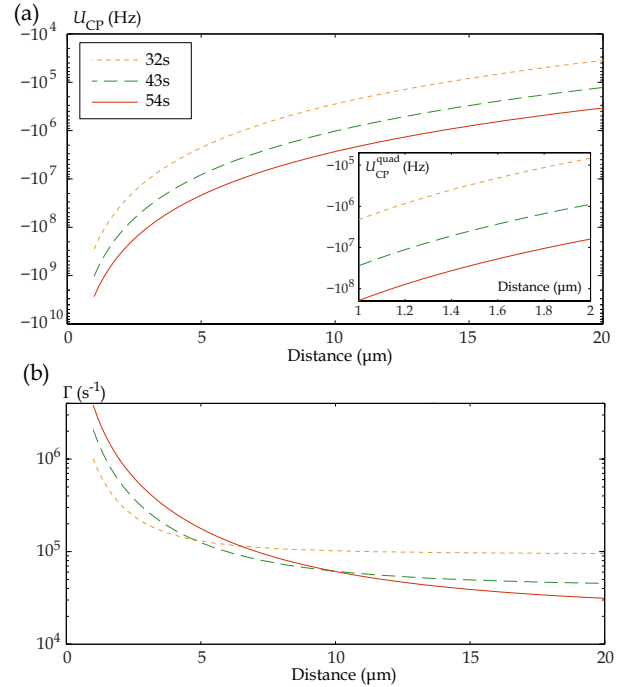}
\caption{(a) Casimir--Polder level shifts of the states $32s$ (dotted line),
$43s$ (dashed line) and $54s$ (solid line) of ${}^{87}$Rb near a copper surface
at 300K: total shift and quadrupole contribution alone (inset). (b)
Spontaneous decay rate near a copper surface at 300K for the 
initial states $32s$ (dotted line), $43s$ (dashed line) and $54s$ (solid line)
of ${}^{87}$Rb.}
\label{fig:potential}
\end{figure}

Figure~\ref{fig:potential}a shows the total CP potential (and hence level shifts)
$U_\mathrm{CP}=U_\mathrm{CP}^\mathrm{dip}+U_\mathrm{CP}^\mathrm{quad}$ for
various $ns$ states (with $n=32,43,54$) of ${}^{87}$Rb near a copper surface at 300K. 
As we are not interested in a particular transition channel, the weighted sum over all 
possible final states has been taken.
One observes that for very small (yet experimentally achievable and indeed
desirable) distances of less than $2\mu$m the expected level shifts rapidly grow
to GHz sizes. At these distances, we also observe significant deviations of the
total shift from the dipole contribution (\ref{cpd}) alone due to the
increasingly important quadrupole shifts [Eq.~(\ref{cpq})] which themselves can
be as large as several MHz (inset in Fig.~\ref{fig:potential}a).

Related to the energy level shift is a line broadening effect, i.e. an increased
rate of spontaneous decay due to strong non-radiative processes, as the atom
approaches the surface \cite{slovaca,decay1}. This strongly enhanced
body-induced spontaneous decay partially counteracts the expected increase in
lifetime as a function of the principal quantum number $n$ in free space
($\Gamma_0\propto n^{-3}$) \cite{gallagher}.

In Fig.~\ref{fig:potential}b we show the total decay rates of the Rydberg states
$ns$ ($n=32,43,54$) of ${}^{87}$Rb as a function of atom-surface distance.
The body-induced decay rates for electric dipole and quadrupole transitions are
calculated from Ref.~\cite{decay1} as
[$\Gamma_{nk}=\Gamma_{nk}^\mathrm{dip}+\Gamma_{nk}^\mathrm{quad}$]
\begin{gather}
\Gamma_{nk}^\mathrm{dip}(\mathbf{r}_A) =
\frac{\omega^2_{nk}}{\hbar\varepsilon_0c^2}
(\mathbf{d}_{nk} \otimes \mathbf{d}_{kn}) \bullet
\mathrm{Im}\,\bm{G}(\mathbf{r}_A,\mathbf{r}_A,|\omega_{nk}|) \nonumber \\ 
\times \big\{\Theta(\omega_{nk})[n(\omega_{nk})+1]
+\Theta(\omega_{kn})n(\omega_{kn})\big\}\,,\\
\Gamma_{nk}^\mathrm{quad}(\mathbf{r}_A) =
\frac{\omega^2_{nk}}{\hbar\varepsilon_0c^2}
(\mathbf{Q}_{nk} \otimes \mathbf{Q}_{kn}) \nonumber \\
\bullet \left[
\bm{\nabla}\otimes\mathrm{Im}\,\bm{G}(\mathbf{r}_A,\mathbf{r}_A,|\omega_{nk}|)
\otimes\overleftarrow{\bm{\nabla}} \right] \nonumber \\
\times \big\{\Theta(\omega_{nk})[n(\omega_{nk})+1]
+\Theta(\omega_{kn})n(\omega_{kn})\big\}\,.
\end{gather}
Note that, unlike the CP potential, the decay rates are always temperature-dependent. 
One observes a strong increase of the decay rates near the surface
($z\lesssim 10\mu\mbox{m}$) which becomes more pronounced for states with higher
principal quantum number $n$. This translates into a relative line broadening of
more than three orders of magnitude that potentially limits trapping and
manipulation times of high-lying states near surfaces. For larger
distances ($z\gtrsim 15\mu\mbox{m}$) the rates quickly approach their free-space
values and show the expected suppression with increasing $n$.

Finally, we will briefly consider how the CP potential and transition rates
scale with atom-surface distance $z$ and the principal quantum number $n$. For
metal surfaces, the reflection coefficients are nearly independent of $\omega$
at infrared frequencies and below. In the low-temperature limit, when the
thermal photon number is negligible, there is no $\omega$ dependence for either
the CP potential or the transition rates. The dipole and quadrupole moments for
the dominant $ns\rightarrow (n-1)p$ 
%SS
and $ns\rightarrow (n-1)d$ 
transitions scale as $n^2$ and $n^4$, respectively, for large $n$. In the
non-retarded limit (valid for surface distances beyond even 100~$\mu$m), the
body-induced rates and the CP potential scale as $z^{-3}$ and $z^{-5}$ for the
dipole and quadrupole contributions, respectively \cite{decay1}. Combining these
results leads to a scaling behaviour of
\begin{gather}
\left| U^\mathrm{dip}_\mathrm{CP} \right| \,, \Gamma_{nk}^{\mathrm{dip}} 
\propto \frac{n^{4}}{z^{3}}, \qquad
\left| U^\mathrm{quad}_\mathrm{CP} \right| \,, \Gamma_{nk}^{\mathrm{quad}}
\propto \frac{n^{8}}{z^{5}},
\end{gather}
for the dipole and quadrupole components of the CP potential $U_\mathrm{CP}$
and the decay rate $\Gamma_{nk}$. 

In the high-temperature limit, the scaling of the CP shifts remains the same due
to the temperature-independence demonstrated in  Ref.~\cite{tinvariance},
whereas the transition rates become proportional to the mean photon number
$n(\omega)\approx k_BT/(\hbar\omega)$. For the dominant dipole and quadrupole
transitions (Fig.~\ref{fig:transitions}), one finds $\omega\propto n^{-3}$, and
the transition rates scale as $\Gamma_{nk}^{\mathrm{dip}} \propto n^7/z^3$ and
$\Gamma_{nk}^{\mathrm{quad}} \propto n^{11}/z^5$, respectively.

We have shown in this Letter that the interaction between highly excited atoms
and macroscopic surfaces leads to energy level shifts that can be as large as
several GHz. This implies that any scheme that relies on the manipulation of
(trapped) Rydberg atoms near surfaces has to account for this major adjustment.
Moreover, some of the advantages of using highly excited Rydberg atoms, in
particular their rapidly decreasing Einstein coefficients with increasing
principal quantum number $n$, are counteracted by the atom-surface interactions.

We thank D.~Cano and J.~Fort\'{a}gh for helpful discussions.
This work was supported by the UK Engineering and Physical Sciences Research
Council. Support from the European Science Foundation (ESF) within the activity 
`New Trends and Applications of the Casimir Effect'
(\texttt{www.casimir-network.com}) is greatfully acknowledged.

\end{document}